# SNR-Progressive Model with Harmonic Compensation for Low-SNR Speech Enhancement

Zhongshu Hou*, Tong Lei*, Qinwen Hu, Zhanzhong Cao, Ming Tang, and Jing Lu, *Member, IEEE*

*Abstract*—Despite significant progress made in the last decade, deep neural network (DNN) based speech enhancement (SE) still faces the challenge of notable degradation in the quality of recovered speech under low signal-to-noise ratio (SNR) conditions. In this letter, we propose an SNR-progressive speech enhancement model with harmonic compensation for low-SNR SE. Reliable pitch estimation is obtained from the intermediate output, which has the benefit of retaining more speech components than the coarse estimate while possessing a significantly higher SNR than the input noisy speech. An effective harmonic compensation mechanism is introduced for better harmonic recovery. Extensive experiments demonstrate the advantage of our proposed model. A multi-modal speech extraction system based on the proposed backbone model ranks first in the ICASSP 2024 MISP Challenge: https://mispchallenge.github.io/mispchallenge2023/index.html.

## I. INTRODUCTION

WITH the advent and great advances of deep neural network (DNN) [1], modern data-driven speech enhancement (SE) systems have outperformed traditional rule-based signal processing methods, and have been applied widely in telecommunication, robust automatic speech recognition (ASR) [2] and hearing aids [3]. Time-domain DNN-based methods enhance speech waveforms directly [4-6], whereas the majority of SE models operate in the time-frequency (T-F) domain to estimate a mask between clean and degraded spectrum [7,8], or to directly predict the clean complex spectrum [9,10]. Different network configurations have been used for spectral and temporal feature extraction in T-F models, including convolutional neural network (CNN) [11-13], recurrent neural network (RNN) [13-15], and self-attention (SA) mechanisms [16,17].

Despite great advances in DNN-based methods, SE still struggles to accurately retain speech components in low-SNR conditions where the desired speech is severely obscured by noise [18]. To tackle this problem, the coarse-to-refine strategy can be utilized, which consists of an initial coarse estimation network and a subsequent inpainting model [19]. However, speech inpainting itself is a challenging task, especially when important features are lost in the first coarse estimation stage. Generative models can be utilized to enhance the inpainting performance or even directly reconstruct desired speech [20,21], but they face the risk of producing "vocalizing" artifacts with poor articulation and no linguistic meaning in low-SNR conditions [22]. Another exploitable strategy is SNR-progressive learning [23], which decomposes the original low-SNR SE task into a series of intermediate targets, each with a small SNR increment. However, due to the lack of an effective compensation mechanism, this strategy still faces the difficulty of recovering severely corrupted speech. It has been noted that estimating pitch and subsequently enhancing the harmonic structure can help improve the quality of the recovered desired speech [24-26]. However, all the currently available methods estimate pitch from the noisy input, which is highly likely biased in low-SNR conditions since the pitch can be merged by intense noise.

In this letter, we propose an SNR-progressive SE model with harmonic compensation for SE in low-SNR conditions. Instead of estimating pitch directly from the noisy input, we estimate pitch from an intermediate output featuring a notable SNR gain. Moreover, we propose a magnitude-based spectrum compensation method for effective harmonic recovery. By implementing an improved version of the state-of-the-art (SOTA) TF-GridNet [27] block as the SE block, we conduct extensive experiments and validate the advantage of the proposed strategy.

## II. METHODOLOGY

### A. System Overview

Let $\mathbf{S}$, $\mathbf{X} \in \mathbb{C}^{T \times F}$ denote clean and noisy complex spectrograms, where $T$ and $F$ denote the time and frequency dimensions, respectively. To gradually enhance the SNR level of the degraded signal, an SNR-progressive SE model, denoted as $\mathcal{F}_{SPSE}$, is first applied to estimate a series of intermediate outputs with specific SNR gain,

$$\tilde{\mathbf{S}}_1, \tilde{\mathbf{S}}_2, \dots, \tilde{\mathbf{S}}_{K+1} = \mathcal{F}_{SPSE}(\mathbf{X}), \quad (1)$$

where $K$ denotes the number of intermediate outputs and $\tilde{\mathbf{S}}_{K+1}$ represents the final coarse estimation of $\mathbf{S}$. To extract the harmonic information of speech, a pitch filtering module, denoted as $\mathcal{F}_{PF}$, is utilized to capture the harmonic structure from $\mathbf{X}$ based on the intermediate result $\tilde{\mathbf{S}}_K$,

$$\mathbf{X}_{PF} = \mathcal{F}_{PF}(\tilde{\mathbf{S}}_K, \mathbf{X}), \quad (2)$$

where $\mathbf{X}_{PF}$ denotes the filtered spectrogram. A harmonic compensation module, denoted as $\mathcal{F}_{HC}$, reconstructs the harmonics of speech based on $\tilde{\mathbf{S}}_{K+1}$ and $\mathbf{X}_{PF}$,

Manuscript received —, 2024; revised —, 2024; accepted —, 2024. Date of publication —, 2024; date of the current version —, 2024. The National Natural Science Foundation of China supported this work with grant number 12274221. The associate editor coordinating the review of this manuscript and approving it for publication was —. (Corresponding author: Jing Lu)

T. Lei, Z. Hou, Q. Hu and J. Lu are with the Key Laboratory of Modern Acoustics, Nanjing University, Nanjing 210008, China and with the NJU-Horizon Intelligent Audio Lab, Horizon Robotics, Beijing 100094, China (email: tonglei@smail.nju.edu.cn; zhongshu.hou@smail.nju.edu.cn; lujing@nju.edu.cn); Z. Cao is with the Nanjing Institute of Information Technology, Nanjing 210036, China (email: zzcao@189.cn); M. Tang is with the State Grid Jiangsu Electric Power Co. Ltd, Nanjing 210024, China (email: tangming930702@163.com)

Digital Object Identifier —

* These authors contributed equally to this work.

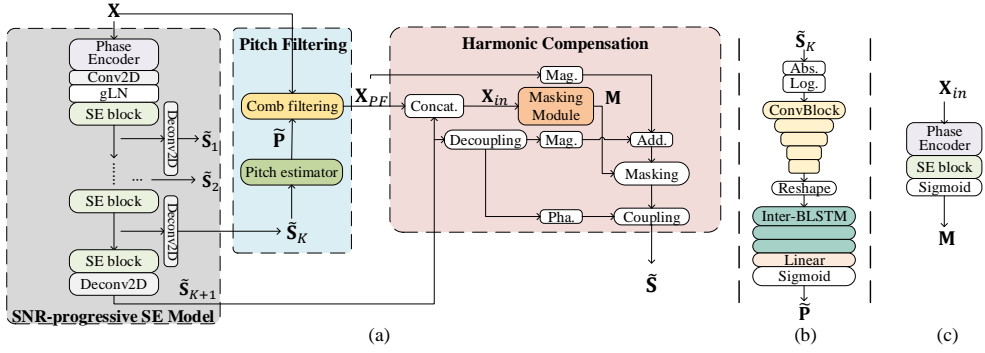

Fig. 1. Diagram of (a) the proposed network, (b) Pitch estimator, (c) Masking Module.

$$\tilde{\mathbf{S}} = \mathcal{F}_{HC}(\tilde{\mathbf{S}}_{K+1}, \mathbf{X}_{PF}), \quad (3)$$

where $\tilde{\mathbf{S}}$ denotes the final enhanced result. The overall architecture of the proposed network is shown in Fig. 1 (a).

*B. SNR-Progressive SE*

SNR-progressive learning decodes the intermediate data stream in DNNs, targeting the intermediate speech with progressively higher SNRs. As shown in Fig. 1 (a), the noisy spectrogram $\mathbf{X}$ is first processed by a phase encoder [28], a 2-dimentional convolution layer (Conv2D), and a global layer normalization layer (gLN) [4] sequentially, and then sent to a series of $K+1$ SE blocks. Between two adjacent SE blocks, a 2-dimentional deconvolution layer (Deconv2D) decodes the network stream and estimates an intermediate target $\mathbf{S}_k, k = 1,2,…,K$, with an increasing SNR level, which can be expressed as

$$\text{SNR}(\mathbf{S}_k, \mathbf{S}) - \text{SNR}(\mathbf{X}, \mathbf{S}) = k\Delta_{SNR}, \quad (4)$$

where $\text{SNR}(*, \mathbf{S})$ calculates the SNR value of given signal compared with the clean target $\mathbf{S}$, and $\Delta_{SNR}$ denotes specific SNR gain in dB. The final output $\tilde{\mathbf{S}}_{K+1}$ is a direct estimation of $\mathbf{S}$.

*C. Pitch Filtering*

For low-SNR SE, model output aimed at noise-free target tends to lose more speech components. In this letter, we propose to estimate the pitch of speech $f_0$ based on the intermediate output $\tilde{\mathbf{S}}_K$, which has the benefit of retaining more speech components than the coarse estimate $\tilde{\mathbf{S}}_{K+1}$ while having a significant higher SNR than the input noisy speech whose pitch may be overwhelmed by intense noise. The pitch filtering module can also be seen in Fig. 1 (a). Similar to [26], the target $f_0$ is obtained from the clean speech with pYIN algorithm [29] and its range is discretized into $N$ target frequencies. An extra dimension is added for unvoiced frames and the Gaussian smoothed $f_0$ label is calculated [26], leading to a target pitch label matrix $\mathbf{P} \in \mathbb{R}^{T\times(N+1)}$. The structure of the pitch estimator is shown in Fig. 1 (b). The logarithmic magnitude spectrum of $\tilde{\mathbf{S}}_K$ is downsampled along frequency dimension by 5 cascaded ConvBlocks, each consisting of a Conv2D, a batch normalization layer (BN), and a PReLU activation function sequentially. The output of ConvBlocks is reshaped to a size $T \times (F' \cdot C')$, where $C'$ and $F'$ denote the channel and compressed frequency dimensions, respectively, and further processed by 3 cascaded bidirectional long short-term memory (BLSTM) layers along time dimension (inter-BLSTM). The output of inter-BLSTMs is fed to a linear layer with a sigmoid activation function, leading to an estimated pitch matrix $\tilde{\mathbf{P}}$. A symmetric non-causal comb filter is used for harmonic enhancement [25],

$$W(Z) = \sum_{i=-1}^{1} \omega_i z^{-i\tau}, \quad (5)$$

where $\tau$ denotes the period of $f_0$ and $\omega_i$ is the normalized Hanning window coefficient. Comb filtering is implemented by using Conv2D for faster inference [26] and applied on the noisy input $\mathbf{X}$, yielding the filtered spectrogram $\mathbf{X}_{PF}$.

*D. Harmonic Compensation*

Compensation or refinement model is commonly used in multi-stage SE network. Complex addition & masking method entails the addition of the compensation spectrum to the coarse estimation, followed by the application of a complex ratio mask (CRM) [26]. In contrast, the complex masking & addition method indicates that the masking operation is executed before the complex addition step [10,24]. In this letter, we propose another compensation scheme for harmonic enhancement, as shown in the area with pink background of Fig. 1 (a). The magnitude of the filtered spectrogram $\mathbf{X}_{PF}$ is first added to the magnitude of coarse estimation $\tilde{\mathbf{S}}_{K+1}$ and then masked by a spectral magnitude mask (SMM) [30]. The phase information of the coarse estimate is further utilized to obtain the final enhanced result $\tilde{\mathbf{S}}$. The operation can be expressed as

$$\mathbf{X}_{in} = \text{Concat}(\tilde{\mathbf{S}}_{K+1}, \mathbf{X}_{PF}), \quad (6)$$
$$\mathbf{M} = \mathcal{F}_M(\mathbf{X}_{in}), \quad (7)$$
$$\tilde{\mathbf{S}}_{mag} = \mathbf{M} \odot (|\tilde{\mathbf{S}}_{K+1}| + |\mathbf{X}_{PF}|), \quad (8)$$
$$\tilde{\mathbf{S}}_{pha} = \text{Phase}(\tilde{\mathbf{S}}_{K+1}), \quad (9)$$
$$\tilde{\mathbf{S}} = \{\tilde{\mathbf{S}}_{mag}, \tilde{\mathbf{S}}_{pha}\}, \quad (10)$$

where operator $\text{Concat}(*,*)$ denotes concatenation along the channel dimension, $\mathcal{F}_M$ denotes the masking module whose structure is shown in Fig. 1 (c), $\mathbf{M} \in \mathbb{R}^{T\times F}$ denotes the SMM output, $\odot$ denotes element-wise multiplication, and $\text{Phase}(*)$ indicates acquiring the phase angle of the complex spectrum.

*E. Training Strategy and Loss Functions*

The training procedure can be divided into three stages. In the first stage, the SNR-progressive SE model is trained with the spectral power compress loss function and the SNR loss function,

$$\mathcal{L}_{RI}(\tilde{\mathbf{S}}, \mathbf{S}) = \|\mathbf{S}_{real}^{\mathcal{C}} - \tilde{\mathbf{S}}_{real}^{\mathcal{C}}\|_F^2 + \|\mathbf{S}_{imag}^{\mathcal{C}} - \tilde{\mathbf{S}}_{imag}^{\mathcal{C}}\|_F^2, \quad (11)$$
$$\mathcal{L}_{Mag}(\tilde{\mathbf{S}}, \mathbf{S}) = \||\mathbf{S}|^\gamma - |\tilde{\mathbf{S}}|^\gamma\|_F^2, \quad (12)$$
$$\mathcal{L}_{Freq}(\tilde{\mathbf{S}}, \mathbf{S}) = \alpha \mathcal{L}_{Mag}(\tilde{\mathbf{S}}, \mathbf{S}) + \beta \mathcal{L}_{RI}(\tilde{\mathbf{S}}, \mathbf{S}), \quad (13)$$
$$\mathbf{S}_{real}^{\mathcal{C}} = |\mathbf{S}|^\gamma \cos\theta_S, \mathbf{S}_{imag}^{\mathcal{C}} = |\mathbf{S}|^\gamma \sin\theta_S, \quad (14)$$
$$\mathcal{L}_{Temp}(\tilde{\mathbf{s}}, \mathbf{s}) = 0.5\sum_t\{\log10[(\mathbf{s}(t) - \tilde{\mathbf{s}}(t))^2]$$
$$- \log10[\mathbf{s}(t)^2]\}, \quad (15)$$

$$\mathcal{L}_{Ovrl}(\tilde{\mathbf{S}}, \mathbf{S}, \tilde{\mathbf{s}}, \mathbf{s}) = \mathcal{L}_{Freq}(\tilde{\mathbf{S}}, \mathbf{S}) + \lambda \mathcal{L}_{Temp}(\tilde{\mathbf{s}}, \mathbf{s}), \quad (16)$$

where $\gamma$ refers to the compression parameter, $\alpha = 0.7$ and $\beta = 0.3$ refer to weighting parameters, the superscript $\mathcal{C}$ denotes the power compressed pattern, $\boldsymbol{\theta}_S$ denotes the phase angle of complex spectrogram, $||*||_F$ refers to the Frobenius norm of the matrix, $\tilde{\mathbf{s}}$ and $\mathbf{s}$ are temporal waveform vectors of $\tilde{\mathbf{S}}$ and $\mathbf{S}$, and $\lambda = 1$ denotes a trade-off parameter. For progressive learning, the overall loss function is

$$\mathcal{L}_{PL} = \sum_{k=1}^{K+1} \mathcal{L}_{Ovrl}(\tilde{\mathbf{S}}_k, \mathbf{S}_k, \tilde{\mathbf{s}}_k, \mathbf{s}_k). \quad (17)$$

In the second stage, pitch estimator is trained together with the SNR-progressive SE model, where only the parameters of pitch estimator are learnable. The binary cross entropy (BCE) is utilized for training,

$$\mathcal{L}_{pitch} = -\sum_{i,j} \{\mathbf{P}(i,j) \log \tilde{\mathbf{P}}(i,j) + [\mathbf{1} - \mathbf{P}(i,j)] \log [1 - \tilde{\mathbf{P}}(i,j)]\}, \quad (18)$$

where $i$ and $j$ denote the time and frequency indexes in the pitch matrix.

In the eventual harmonic compensation stage, the whole network is trained together, where only the parameters of masking module and the last SE block in SNR-progressive SE model are learnable. The loss function can be expressed as

$$\mathcal{L}_{HC} = \mathcal{L}_{Ovrl}(\tilde{\mathbf{S}}, \mathbf{S}, \tilde{\mathbf{s}}, \mathbf{s}). \quad (19)$$

### III. EXPERIMENTS

#### A. SE Block and Datasets

Compared with previous progressive SE models comprising simple cascaded linear or RNN layers [23], we choose the SOTA TF-GridNet [27] as SE model, where a series of TF-GridNet blocks are implemented as SE blocks in the proposed model. For better phase modeling, an effective phase encoder module [28] is utilized to map complex spectral features to real. We verify the effectiveness of our proposed method on the speech dataset of the DNS-3 challenge [31] and noise data from both the DNS-3 and DCASE challenge [32]. We generate 1000 hours of noisy-clean pairs. During mixing process, room impulse responses (RIR) are convolved with clean speech to generate reverberant speech, the early reflection (first 50ms) of which is preserved as the final training target. The intermediate targets are generated by mixing reverberant speech and noise with given SNR levels. In the training stage, reverberant utterances are mixed with noise signals with SNRs ranging from −15 dB to 0 dB. For the test set, 491 clips of reverberant utterances are mixed with unseen noise with SNRs ranging from −15 dB to 15 dB. Note the test noise does not include speech-like noise like baby cries. which would potentially affect the accuracy of the harmonic compensation module (HC). All audio clips are sampled at 16 kHz and each test clip is 10 s long.

#### B. Parameter Setup and Training Configuration

The window size and hop length of short-time Fourier transformation (STFT) are 32ms and 16ms, respectively. The discrete Fourier transformation length is 512 and a periodic Hanning widow is used for overlap-add waveform reconstruction. For phase encoder, the output channel is set to 4. The kernel size and stride of time and frequency dimensions are (3, 1) and (1, 1), respectively. In the SNR-progressive SE model, the SNR gain $\Delta_{SNR}$ is set to 5 dB. The kernel sizes and strides of Conv2D and all Deconv2Ds are (3, 3) and (1, 1), respectively, and 5 TF-GridNet[1] blocks are utilized, i.e. $K = 4$. For each TF-GridNet block, the embedding dimension for each T-F unit is set to 32. The number of hidden units of each BLSTM layer is set to 100 and the number of heads in self-attention is 4. The unfolding layer uses a kernel size of 4 and stride size of 1. The network details of phase encoder and TF-GridNet block can be found in [28] and [27], respectively. For pitch estimator, the output channel of Conv2D is set to [16, 32, 64, 128, 256] in each ConvBlock. The kernel sizes and strides of all Conv2Ds are set to (3, 3) and (1, 2), respectively. The number of hidden units is set to [512, 256, 128] in each BLSTM layer. The target $f_0$ range is set as [62.5, 500] Hz and the $N$ is 225. A TF-GridNet block is chosen as the SE block in the masking module and the parameter setup is the same as that in the SNR-progressive SE model.

The batch size in our training is 8 and the temporal length of input audio is 8s. Warmup strategy [16] is critical in training self-attention based model, where the learning rate $\nu$ is updated with the rule: $\nu = \min(1/\sqrt{\varphi}, \varphi/\sqrt{\Psi^3})/\sqrt{C}$, with $C = 100$, warmup steps $\Psi = 10000$ and $\varphi$ denoting the training step. We train the model by the warmup-based Adam optimizer with $\beta_1 = 0.9$, $\beta_2 = 0.98$, and $\epsilon = 10^{-9}$. The compression parameter $\gamma$ is 1/3. The number of training steps for each epoch is 1250 and the training period is 150 epochs in total.

#### C. Ablation Study and Evaluation Metrics

We conduct experiments on the proposed model and the original TF-GridNet [27]. For a fair comparison, TF-GridNet is implemented with 6 TF-GridNet blocks with aforementioned parameter setup to ensure almost equivalent model complexity to the proposed network. MTFAA [28], ranked 1st place in DNS-4 challenge, and CMGAN [33], ranked 1st place in the public VCTK-Demand dataset, are also tested as SOTA baseline models. We investigate the contribution of each module of our proposed network by removing specific modules in ablation tests.

Objective metrics are used to evaluate the performance, i.e., perceptual evaluation of speech quality (PESQ) [34], short-time objective intelligibility (STOI) [35], signal distortion ratio (SDR), and composite mean opinion score (MOS) based metrics [36] with MOS prediction of the signal distortion (CSIG), MOS prediction of the intrusiveness of background noise (CBAK) and MOS prediction of the overall effect (COVL). Higher values indicate better performance for all metrics. Average scores over SNR ranges are given to clearly present the overall performance gap between different models.

#### D. Experimental Results and Analysis

*1) Comparison with the State-of-the-Art Models*

It can be seen from Table I that the proposed model outperforms all previous SOTA models in terms of all objective metrics. In low-SNR scenarios (−15 dB ~ −5 dB), STOI is a more critical metric since it reflects whether the speech information can be understood. Our system achieves a more considerable gain on STOI compared with MTFAA and CMGAN (0.288 vs. 0.168 and 0.190). Note that our proposed method shows a more significant STOI increment in low-SNR conditions compared to higher-SNR conditions (0.288 vs. 0.224 and 0.097). With similar model complexity, our proposed method achieves a better

---

1. The source code of TF-GridNet can be found in
https://github.com/espnet/espnet/blob/master/espnet2/enh/separator/tfgridnet_separator.py.

TABLE I
COMPARISON OF SE PERFORMANCE WITH DIFFERENT MODELS

| Metrics | | PESQ | | | | STOI | | | | SDR (dB) | | | |
|---|---|---|---|---|---|---|---|---|---|---|---|---|---|
| SNR (dB) | Para. (M) | $-15 \sim -5$ | $-5 \sim 5$ | $5 \sim 15$ | Avg. | $-15 \sim -5$ | $-5 \sim 5$ | $5 \sim 15$ | Avg. | $-15 \sim -5$ | $-5 \sim 5$ | $5 \sim 15$ | Avg. |
| Noisy | | 1.089 | 1.098 | 1.313 | 1.170 | 0.478 | 0.655 | 0.832 | 0.655 | -10.06 | -0.389 | 8.327 | -0.772 |
| MTFAA | 2.22 | 1.274 | 1.654 | 2.078 | 1.669 | 0.646 | 0.799 | 0.880 | 0.775 | 1.045 | 6.839 | 9.887 | 5.924 |
| CMGAN | 1.83 | 1.334 | 1.776 | 2.248 | 1.786 | 0.668 | 0.822 | 0.896 | 0.795 | 2.257 | 8.015 | 10.409 | 6.894 |
| TF-GridNet | 2.75 | 1.413 | 1.982 | 2.400 | 1.932 | 0.737 | 0.871 | 0.924 | 0.844 | 4.107 | 9.860 | 12.672 | 8.880 |
| **Proposed** | 2.79 | **1.584** | **2.076** | **2.511** | **2.057** | **0.766** | **0.879** | **0.929** | **0.858** | **4.218** | **10.24** | **13.430** | **9.296** |
| $-$PE | 2.78 | 1.462 | 2.009 | 2.424 | 1.965 | 0.728 | 0.873 | 0.922 | 0.841 | 4.139 | 9.613 | 13.137 | 8.963 |
| $-$HC | 2.31 | 1.471 | 2.018 | 2.439 | 1.976 | 0.725 | 0.872 | 0.919 | 0.839 | 4.131 | 9.562 | 13.112 | 8.935 |
| $-$HC$-$PL | 2.31 | 1.381 | 1.921 | 2.327 | 1.876 | 0.712 | 0.863 | 0.910 | 0.828 | 3.973 | 9.198 | 12.486 | 8.552 |
| Metrics | | CSIG | | | | CBAK | | | | COVL | | | |
| SNR (dB) | Para. (M) | $-15 \sim -5$ | $-5 \sim 5$ | $5 \sim 15$ | Avg. | $-15 \sim -5$ | $-5 \sim 5$ | $5 \sim 15$ | Avg. | $-15 \sim -5$ | $-5 \sim 5$ | $5 \sim 15$ | Avg. |
| Noisy | | 1.659 | 1.965 | 1.864 | 1.830 | 1.493 | 1.696 | 1.619 | 1.602 | 1.305 | 1.503 | 1.437 | 1.415 |
| MTFAA | 2.22 | 2.820 | 2.789 | 2.767 | 2.792 | 2.380 | 2.201 | 2.223 | 2.268 | 2.218 | 2.158 | 2.072 | 2.149 |
| CMGAN | 1.83 | 2.918 | 2.992 | 2.989 | 2.966 | 2.458 | 2.435 | 2.448 | 2.447 | 2.293 | 2.343 | 2.345 | 2.327 |
| TF-GridNet | 2.75 | 3.141 | 3.275 | 3.286 | 3.234 | 2.685 | 2.718 | 2.732 | 2.711 | 2.498 | 2.592 | 2.623 | 2.571 |
| **Proposed** | 2.79 | **3.211** | **3.298** | **3.301** | **3.270** | **2.803** | **2.768** | **2.779** | **2.783** | **2.601** | **2.632** | **2.658** | **2.630** |
| $-$PE | 2.78 | 3.175 | 3.234 | 3.190 | 3.200 | 2.751 | 2.717 | 2.720 | 2.729 | 2.551 | 2.576 | 2.589 | 2.572 |
| $-$HC | 2.31 | 3.166 | 3.230 | 3.206 | 3.201 | 2.750 | 2.716 | 2.720 | 2.729 | 2.542 | 2.589 | 2.592 | 2.574 |
| $-$HC$-$PL | 2.31 | 3.082 | 3.184 | 3.137 | 3.134 | 2.664 | 2.692 | 2.679 | 2.678 | 2.462 | 2.532 | 2.529 | 2.508 |

$-$ indicates removing this module and **bold** indicates the best score in each case. HC, PE, PL denote harmonic compensation module, phase encoder, and SNR-progressive learning strategy, respectively. Note that removing HC indicates that $\tilde{\mathbf{S}}_{K+1}$ is utilized for evaluation.

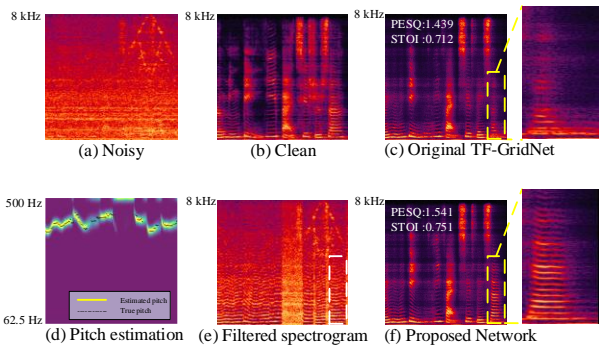

Fig. 2. Typical spectrograms demonstrating the harmonic recovery ability of the proposed network.

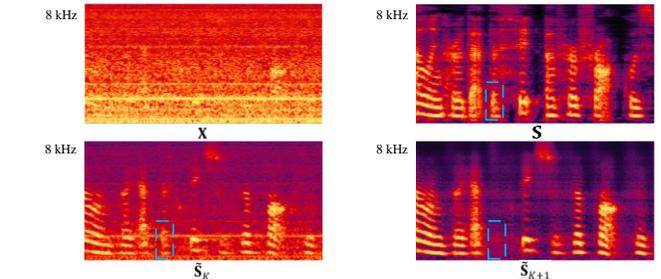

Fig. 3. Typical spectrograms illustrating the rationale underlying employing $\tilde{\mathbf{S}}_K$ for pitch estimation.

TABLE II
PITCH ESTIMATION RESULTS ON TEST SET

| | Accuracy (%) | | |
|---|---|---|---|
| SNR (dB) | $\mathbf{X}$ | $\tilde{\mathbf{S}}_K$ | $\tilde{\mathbf{S}}_{K+1}$ |
| 15 | 98.8 | 98.2 | **99.2** |
| 5 | 93.9 | **97.1** | 97.1 |
| $-5$ | 87.2 | **94.3** | 92.6 |
| $-15$ | 68.1 | **89.9** | 84.0 |

performance than the original TF-GridNet. It can be seen from the typical sample shown in Fig. 2 that the pitch filtering module facilitates harmonic extraction, as shown in the white boxes in Fig. 2 (e). The harmonic compensation module can then recover more harmonic components compared with the original TF-GridNet, as shown in the yellow boxes in Fig. 2 (c) and (f), leading to better speech quality and intelligibility, which can be reflected in the PESQ and STOI metrics in the two figures.

*2) Ablation Study*

The bottom 3 rows in Table I show the ablation results by removing specific modules. It can be seen that removing phase encoder (PE) hampers the overall SE performance, indicating that mapping complex features to a specific dynamic range can help improve the performance of TF-GridNet. Compared with removing PE, removing HC individually results in more STOI drop in low-SNR conditions, demonstrating its efficacy in tackling low-SNR SE problem. When HC and SNR-progressive learning (PL) are removed simultaneously, the model simplifies to a 5-block TF-GridNet, whose performance degrades considerably and becomes inferior to the baseline 6-block TF-GridNet. This validates the necessity of applying SNR-progressive learning in low-SNR tasks.

*3) Accuracy of Pitch Estimation*

To verify the advantage of the proposed pitch filtering strategy, we also conduct experiments with different inputs, i.e., $\mathbf{X}$ (noisy input), $\tilde{\mathbf{S}}_K$ (proposed) and $\tilde{\mathbf{S}}_{K+1}$ (coarse estimate), to the pitch estimator. For each input, the pitch estimator is trained separately. The accuracy of pitch estimation is calculated from the generated test set with different SNR levels, which are presented in Table II. It can be seen that the pitch estimation is highly reliable in high-SNR conditions. However, in low-SNR conditions, the accuracy of estimation from $\tilde{\mathbf{S}}_K$ is significantly higher than that from the other two inputs (89.9% vs. 68.1% and 84.0% at $-15$ dB SNR level), which lays the foundation for a better SE performance. A typical sample depicted in Fig. 3 indicates that $\tilde{\mathbf{S}}_{K+1}$ loses more speech components (shown in the blue boxes) compared with $\tilde{\mathbf{S}}_K$ when $\mathbf{X}$ is contaminated by intense noise, leading to more biased pitch estimation.

IV. CONCLUSION

In this letter, we propose an SNR-progressive speech enhancement model with harmonic compensation. The pitch is estimated from an intermediate output of progressive learning, demonstrating superior accuracy compared to estimations inferred from the noisy input and the final output in low-SNR conditions. A magnitude-based masking approach is proposed to exploit the pitch better and fulfill the harmonic compensation task. The effectiveness of the proposed model is validated through experiments, using the SOTA TF-GridNet as the speech enhancement block. While our model shows promising results, we acknowledge its limitations in complex scenarios with multiple speakers or speech-like interference. Future work may integrate speech separation techniques to address these challenges.